\documentclass[preprint,aps,preprintnumbers,showpacs,nofootinbib,floatfix]{revtex4}
\usepackage{amsmath,amssymb,bm,graphicx}

\renewcommand\L{\mathcal{L}}
\renewcommand\d{\partial}
\newcommand\x{\mathbf{x}}
\renewcommand\v{\mathbf{v}}
\newcommand\p{\mathbf{p}}
\renewcommand\k{\mathbf{k}}
\newcommand\grad{\bm{\nabla}}
\begin{document}

\preprint{INT-PUB 05-06}

\title{Effective Lagrangian and Topological Interactions in Supersolids}
\author{D.\,T.~Son}
\affiliation{Institute for Nuclear Theory,
University of Washington, Seattle, Washington 98195-1550}

\begin{abstract}

We construct a low-energy effective Lagrangian describing
zero-temperature supersolids.  Galilean invariance imposes strict
constraints on the form of the effective Lagrangian.  We identify a
topological term in the Lagrangian that couples superfluid and
crystalline modes.  For small superfluid fractions this interaction
term is dominant in problems involving defects.  As an illustration,
we compute the differential cross section of scatterings of low-energy
transverse elastic phonons by a superfluid vortex.  The result is
model-independent.

\end{abstract}
\date{January 2005}

\maketitle

\emph{Introduction.}---The possibility of superfluid behavior in
solids was considered by Andreev and Lifshitz (AL) in a seminal paper
of 1969~\cite{AndreevLifshitz}.  They proposed that point defects in a
$^4$He crystal may become quantum at low temperatures and form a Bose
condensate.  The possibility of superfluid solids (``supersolids'')
was also conjectured in Refs.~\cite{Chester,Legett}.  For a long
time, the experimental search for a supersolid phase of helium was
unsuccessful.  Recently, however, Kim and Chan claimed that superfluid
behavior may have been observed in solid
$^4$He~\cite{KimChan-Nature,KimChan-Science}.  This has stimulated
renewed interest in the supersolid phase.  The superfluid fraction was
found in Ref.~\cite{KimChan-Science} to be of order $10^{-2}$,
considerably larger than what was theoretically expected
($\lesssim10^{-4}$~\cite{Legett}).

In the following, we assume that the supersolid phase does indeed
occur.  We shall not discuss the microscopic origin of such a
behavior.  Rather, we concentrate our attention on the theoretical
description of the low-energy dynamics of the supersolid phase.  One
can expect that, regardless of the details of the microscopic
mechanism underlying the supersolid state, its low-energy dynamics is
simple and universal, and is dictated by symmetry principles alone.
This is because the low-energy degrees of freedom of supersolids are
the Nambu--Goldstone bosons arising from spontaneous breaking of
translational symmetry and the U(1) symmetry generated by the
conserved particle number.  In fact, Andreev and Lifshitz have
constructed the hydrodynamic equations describing a supersolid (in the
limit of small strain), based on conservation laws and Galilean
invariance only.

In this Letter, we restrict ourselves to zero temperature and derive an
alternative description of supersolids based on a low-energy effective
Lagrangian~\cite{Weinberg:1978kz}.  Such a description is possible
since dissipative effects disappear at $T=0$.  The effective
Lagrangian description holds several advantages over the one based on
the hydrodynamic equations.  As it is generally true, the Lagrangian
formulation enables straightforward application of field-theoretical
techniques such as Feynman diagrams.  In our particular case, the
Lagrangian also elucidates the appearance of a certain ``topological''
interaction term, which is important in processes involving defects.

The form of the effective Lagrangian is constrained by various
symmetries, among which an important role is played by the Galilean
invariance.  For liquid superfluids, the most general
Galilean-invariant effective Lagrangian was constructed by Greiter,
Wilczek, and Witten~\cite{Greiter:1989qb}.  For supersolids the
possible structures appearing in the Lagrangian is richer than in
supersolid $^4$He, but are still rather restrictive.
%Using the Lagrangian we shall compute the differential cross section
%of scattering of elastic waves by a superfluid vortex.
\vspace{6pt}

\emph{Degrees of freedom.}---The fields appearing in the effective
Lagrangian are the four Goldstone bosons which appear due to the
spontaneous breaking of the U(1) particle number symmetry and the
translational symmetry along three spatial directions.  One of these
fields is therefore the phase of the superfluid condensate, $\theta$;
under the action of the particle number, it transforms as
$\theta\to\theta+\alpha$.

The three remaining fields are translation-breaking scalars $X^a$,
$a=1,2,3$, which can be introduced as follows~\cite{Leutwyler:1996er}.
Imagine a system of coordinates $X^a$, $a=1,2,3$ which is frozen in
the body of the solid.  (The choice of the system is completely
arbitrary, but for simplicity one can choose it to coincide with the
coordinates in our reference frame $x^i$, $i=1,2,3$ if the solid is at
equilibrium at some arbitrarily chosen external pressure $P_0$.)  When
the solid moves, this system of coordinates also moves, so if one
follows one particular material point in the solid, its coordinates in
the $X$ system remain constant.  In general, the coordinate system $X$
is curved.  The time history of the solid is completely characterized
by three functions $X^a(t,\x)$, which give the coordinates, in the
comoving frame, of the material point that is located at the position
$\x$ at time $t$.  $X^a(t,\x)$ are the fields that enter the effective
theory together with the U(1) phase $\theta$.

One can expand the fields around the ground state 
\begin{equation}
  \theta = \mu_0 t - \varphi,\qquad X^a = x^a - u^a,
\end{equation} 
where $\varphi$ and $u^a$ fluctuate around zero.  Here $\mu_0$ is the
chemical potential at pressure $P_0$; $u^a$ is the usual displacement
vector.  The superfluid velocity and the strain of the crystal are
related to the first spatial derivatives of $\theta$ and $X^a$:
\begin{equation}
  \v_s = \frac1m \grad\varphi,\qquad
  \d_i X^a = \delta_{ia} - \d_i u^a
\end{equation}
(in most of this Letter, $\hbar=1$).

The density of lattice sites is a constant in $X$ space.  We denote
this constant by $n_0$.  In the $x$ space, the density of lattice sites
is
\begin{equation}
  n_0 \det |\d_i X^a| = \frac{n_0}6\epsilon_{ijk}\epsilon_{abc} 
  \d_i X^a \d_j X^b \d_k X^c .
\end{equation}
For an ordinary crystal, nonsuperfluid and without defects, this
coincides with the particle number density.
\vspace{6pt}

\emph{Derivative expansion.}---The effective Lagrangian should be
invariant under the U(1) particle number symmetry,
$\theta\to\theta+\alpha$, and coordinate shift in the frozen frame,
$X^a\to X^a+\alpha^a$.  Thus the Lagrangian contains only time and
coordinate derivatives of $\theta$ and $X^a$, but not $\theta$ or
$X^a$ by themselves.

In order to discuss the low-energy regime, we follow the standard
effective field theory philosophy and perform a derivative expansion
of the effective Lagrangian.  Because $\theta$ and $X^a$ do not appear
without derivatives, there are two different ways to perform the
derivative expansion.  One possibility is to assume that $\theta$ and
$X^a$ are slowly varying functions of space and time.  Alternatively,
one can assume that the first (temporal and spatial)
\emph{derivatives} of $\theta$ and $X^a$ vary slowly.  Clearly, the
second alternative is more general, since it allows for the first
derivatives of $\theta$ and $X^a$ to be large.  In particular, the
superfluid velocity and the strain of the crystal do not have to be
small.  In our subsequent discussion, we will therefore assume that
$\dot\theta$, $\d_i\theta$, $\dot X^a$ and $\d_i X^a$ may be not
small, but vary slowly in space and time over distances large compared
to all microscopic length scales, such as the superfluid healing
length.

Keeping only leading-order terms in the derivative expansion, the
Lagrangian is a function of the first derivatives of fields,
\begin{equation}\label{L-norot}
  \L = \L(\dot\theta,\, \d_i\theta,\, \dot X^a,\, \d_i X^a).
\end{equation}
This Lagrangian, in general, contains terms to all orders of fields.
In each term in the series expansion over fields $\theta$ and $X^a$,
one keeps the lowest possible number of derivatives equal to the
number of fields.
\vspace{6pt}

\emph{Rotational invariance.}---The effective Lagrangian should be
invariant under spatial rotations.  The fields $X^a$, despite being a
three-component field, transform under spatial rotations like
\emph{scalar} fields.  This is because $X^a$ are the coordinates of
the internal system frozen in the solid body, which are not rotated
with the axes of spatial coordinates $x^i$.  Therefore the Lagrangian
(\ref{L-norot}) is a function of the following rotationally invariant
combinations of arguments,
\begin{equation}\label{L-rot}
  \L = \L(\dot\theta,\, \dot X^a,\, 
       (\d_i\theta)^2,\, \d_i\theta\d_i X^a,\, u^{ab})\,,
\end{equation}
where we introduce the notation
\begin{equation}\label{uab}
  u^{ab} = \d_i X^a \d_i X^b = \delta^{ab} - \d_a u^b - \d_b u^a
    + \d_i u^a \d_i u^b .
\end{equation}
It can be shown that any rotationally invariant function of first
derivatives of $\theta$ and $X^a$ can be written as a function of the
parameters staying in Eq.~(\ref{L-rot}).  In particular,
\begin{eqnarray}
  & & \epsilon_{ijk}\d_i X^a \d_j X^b \d_k X^c 
      =  \epsilon^{abc} \sqrt{\det u} \label{iden1}\,,\\
  & & \epsilon_{ijk}\d_i\theta\d_j X^a \d_k X^b
      = \epsilon^{abc} \sqrt{\det u}\,u^{-1}_{cd}\d_i\theta\d_i X^d ,
      \label{iden2}
\end{eqnarray}
where $\det u$ is the determinant of the $3\times3$ matrix $u^{ab}$,
and $u^{-1}_{ab}$ is the inverse matrix of $u$:
$u^{-1}_{ab}u_{bc}=\delta_{ac}$.  
\vspace{6pt}

\emph{Galilean invariance.}---Further constraints on the effective
Lagrangian follow from Galilean invariance.  In a nonrelativistic
theory where all particles have the same mass, the momentum density is
proportional to the particle number flux,
\begin{equation}\label{Galilean}
  T^{0i} = m j^i ,
\end{equation}
where $m$ is the mass of the $^4$He atom.  The momentum density and
the particle number flux are found from the Lagrangian by using
Noether's theorem,
\begin{equation}
  T^{0i} = -\frac{\d\L}{\d\dot\theta}\d_i\theta 
          -\frac{\d\L}{\d\dot X^a}\d_i X^a, \qquad
  j^i = \frac{\d\L}{\d(\d_i\theta)}\,.
\end{equation}
The most general form of the Lagrangian which is consistent with
rotational [Eq.~(\ref{L-rot})] and Galilean invariance
[Eq.~(\ref{Galilean})] is
\begin{equation}\label{L-Gal}
  \L = \L(\mu, w^a, u^{ab})\,.
\end{equation}
Here $u^{ab}$ was defined in Eq.~(\ref{uab}), and $\mu$ and $w^a$ are
\begin{equation}\label{muw}
  \mu = \dot\theta - \frac{(\d_i\theta)^2}{2m}\,, \quad
  w^a = -\dot X^a + \frac1m \d_i\theta\d_i X^a .
\end{equation}
The variable $\mu$ appears in the Lagrangian treatment of
superfluids~\cite{Greiter:1989qb}: the crystal structure is absent
there and the Lagrangian is a function of $\mu$ alone.  Physically,
$\mu$ is the local chemical potential as measured in the frame moving
with the superfluid velocity.  The meaning of $w^a$ can be made clear
by expanding it,
\begin{equation}
  \mathbf{w} = \dot {\bf u} + (\v_s\cdot\grad)\mathbf{u} - \mathbf{v}_s \,.
\end{equation}
At the linearized level $w^a$ is the difference between the velocity
of motion of the crystal lattice and the superfluid velocity.
\vspace{6pt}

\emph{Connection to the AL hydrodynamic theory.}---The
zero-temperature AL hydrodynamic equations can be derived from the
Lagrangian~(\ref{L-Gal}) as the equations of conservation of particle
number, energy, and momentum, written using a particular set of
variables.  We introduce the total density $\rho$, the superfluid
velocity $\v_s$, and the vector $\v_n$ which corresponds to the normal
velocity in the AL theory,
\begin{equation}
  \rho = m\frac{\d\L}{\d\mu}\,, \quad v_{si} = \frac1m \d_i\varphi,
  \quad v_{ni} = \frac{\d x^i}{\d X^a} \dot u^a .
\end{equation}
Actually $\v_n$ is the velocity of the crystal lattice; at the
linearized level $\v_n=\dot{\bf u}$.  At zero temperature we do not
have a normal component distinct from the crystal lattice.
Furthermore, we introduce the momentum density $\p$ and the energy
density $\varepsilon$ in the frame $\v_s=0$,
\begin{equation}
  p_i = \frac{\d\L}{\d w^a}\d_i X^a\,,\quad
  \varepsilon = \mu\frac{\d\L}{\d\mu} + w^a\frac{\d\L}{\d w^a} - \L\,,
\end{equation}
and an auxiliary tensor $\lambda_{ia}$,
\begin{equation}\label{lambdaia}
  \lambda_{ia} = 2\frac{\d\L}{\d u^{ab}}\d_i X^b 
  + (v_{ni}-v_{si})\frac{\d\L}{\d w^a}\,.
\end{equation}
After some algebra, we find that the momentum density can be written
as
\begin{equation}\label{j-AL}
  T^{0i} = mj^i = \rho v_{si} + p_i\,,
\end{equation}
the energy density and the energy flux as~\footnote{Note that
$T^{0i}\neq T^{i0}$, since in our nonrelativistic theory energy does
not include rest mass.  Our normalization of the chemical
potential differs from that of AL by a factor of $m$.}
\begin{eqnarray}
  T^{00} &=& \frac{\rho v_s^2}2 + \p\cdot\v_s + \varepsilon,
  \label{T00-AL}\\
  T^{i0} &=& \biggl(\mu + \frac{mv_s^2}2\biggr) j^i + v_{ni}(\v_n\cdot\p)
             -\lambda_{ia}\dot u^a , \label{T0i-AL}
\end{eqnarray}
and the differential of $\varepsilon$ as
\begin{equation}\label{de-AL}
  d\varepsilon = \frac\mu m d\rho + (\v_n-\v_s)\cdot d\p 
    +\lambda_{ia}d(\d_i u^a)\,.
\end{equation}
Equations (\ref{j-AL})--(\ref{de-AL}) are identical to the
corresponding AL equations at $T=0$.  The stress tensor can be
transformed into the form
\begin{equation}\label{Tik}
\begin{split}
  T^{ik} &= \rho v_{si} v_{sk} + v_{sk}p_i + v_{ni}p_k
% \\  &\quad
 +\delta_{ik}
  \Bigl[\mu\frac\rho m + (\v_n-\v_s)\cdot\p -\varepsilon\Bigr]
%  \\ &\quad
  -\lambda_{ik} +\lambda_{ia}\d_k u^a ,
\end{split}
\end{equation}
which \emph{almost} coincides with the corresponding AL expression.
The only difference is the last term on the right hand side of
Eq.~(\ref{Tik}), which is nonlinear in strain and was neglected in
Ref.~\cite{AndreevLifshitz}.  Moreover, from the definition of
$\lambda_{ia}$, Eq.~(\ref{lambdaia}). one can derive the following
relation:
\begin{equation}
\begin{split}
  & \lambda_{ik} - \lambda_{ki} -\lambda_{ia}\d_k u^a + \lambda_{ka}\d_i u^a
%  \\ & \qquad
  = (v_{ni}-v_{si})p_k - (v_{nk}-v_{sk})p_i \,,
\end{split}
\end{equation}
which coincides, up to the two terms nonlinear in strain on the left
hand side, to an equation \emph{postulated} in
Ref.~\cite{AndreevLifshitz} for $T^{ik}$ to be a symmetric tensor.

As one can see, the effective Lagrangian provides an extremely compact
encoding of the hydrodynamic equations.  Terms nonlinear in strain
which were omitted in the AL theory are fully kept in the Lagrangian.
Moreover, for quantum problems (such as scatterings of phonons) it is
easier to work with the Lagrangian than with the field equations.  We
now show that the Lagrangian contains a special topological term that
is important for scattering off defects.  For this end, we first
discuss the nonsuperfluid limit of the Lagrangian~(\ref{L-Gal}).
\vspace{6pt}

\emph{The nonsuperfluid limit}---The superfluid fraction
$\rho_s/\rho$ of solid helium-4, if nonzero, is much smaller than
one: experiments~\cite{KimChan-Nature,KimChan-Science} indicate a
value of order $10^{-2}$, while theoretical arguments~\cite{Legett}
suggest $\rho_s/\rho\lesssim10^{-4}$.  We shall, therefore,
concentrate on the limit $\rho_s\ll\rho$.  To start, let us discuss
the limit of vanishing superfluid density.

One expects that the nonsuperfluid crystalline state is realized as
some particular limit of the supersolid state, where $\theta$
decouples from the dynamics of $X^a$ fields.  This might seem
nontrivial, since in Eq.~(\ref{L-Gal}) the time derivative of $X^a$
enters the Lagrangian in the combination $w^a$ which involves $\theta$
[Eq.~(\ref{muw})].  However, it is possible to achieve such a
decoupling.  Consider the following Lagrangian:
\begin{equation}\label{L-nons}
  \L = \rho_0 \sqrt{\det u}\,\biggl( \frac12 u^{-1}_{ab} w^a w^b 
       + \frac\mu m\biggr) - V(u^{ab})\,,
\end{equation}
where $\rho_0$ is some constant with the dimension of mass density,
$\mu$ and $w$ are defined in Eq.~(\ref{muw}), and $V$ is an arbitrary
function of the strain $u^{ab}$ consistent with lattice symmetry.
Using the identities~(\ref{iden1}) and (\ref{iden2}), this Lagrangian
can be transformed into the form
\begin{equation}\label{L-nonsuper}
\begin{split}
  \L &= \frac{\rho_0}2 \sqrt{\det u}\, u^{-1}_{ab} \dot X^a \dot X^b
      - V(u^{ab})
%  \\ &\quad
  + \frac{\rho_0}{6m} \epsilon^{\mu\nu\lambda\rho}\epsilon_{abc}
     \d_\mu\theta\d_\nu X^a \d_\lambda X^b \d_\rho X^c .
\end{split}
\end{equation}
The Greek indices in the last term are spacetime indices which run
over $t,x,y,z$; $\epsilon^{\mu\nu\lambda\rho}$ is the completely
antisymmetric tensor defined so that $\epsilon^{0123}=1$.  The phase
$\theta$ appears only in the last term of the Lagrangian, which is a
full derivative; thus $\theta$ decouples from the dynamics.  The last
term, which will be called the ``topological term,'' still plays a
useful role: the particle number current computed from
(\ref{L-nonsuper}) by using Noether's theorem arises completely
from this term,
\begin{equation}\label{jtop}
  j^\mu = \frac{\rho_0}{6m}\epsilon^{\mu\nu\lambda\rho}\epsilon_{abc}
      \d_\nu X^a \d_\lambda X^b \d_\rho X^c ,
\end{equation}
from which we see that $\rho_0$ is the total mass density in
equilibrium where $X^a=x^a$.  The current~(\ref{jtop}) is trivially
conserved.  Note that the first term in Eq.~(\ref{L-nonsuper}) can be
written as $mj^i j^i/(2 j^0)$, which is what one expects for the
kinetic energy from Galilean invariance.  This fact shows that the
construction (\ref{L-nons}) is unique.

Now if one allows the superfluid fraction $\rho_s$ to be nonzero and
small, then there are additional terms proportional to $\rho_s$ added
to the Lagrangian, which makes $\theta$ a dynamical field.  The
topological term continues to be present in the Lagrangian with a
coefficient which differs only slightly from $\rho/(6m)$.  This term
is responsible for low-energy scattering of elastic waves by a
superfluid vortex, as we shall see.
\vspace{6pt}

\emph{Scattering of elastic waves by a superfluid vortex.}---Let us
now use the effective Lagrangian to compute the scattering cross
section of elastic waves by a superfluid vortex.  In the presence of a
vortex $\varphi$ is a multivalued function, and $\v_s\sim\grad\varphi$
is singular at the vortex core.  Due to the multivalued nature of
$\theta$, the topological term is no longer integrated by part to
zero.  Integrating by part, this term can be written as
\begin{equation}
  \L_{\rm top} = -\frac\rho{6m}\epsilon^{\mu\nu\lambda\rho}
     \epsilon^{abc}\d_\mu\d_\nu\varphi X^a\d_\lambda X^b \d_\rho X^c .
\end{equation}
Expanded over small perturbations, this expression contains a
term proportional to ${\bf u}\cdot\dot{\bf v}_s$, which was
identified in Ref.~\cite{ChudnovskyKuklov} in the context of an
Abrikosov vortex in a crystal.
For definiteness, consider a vortex located at $x=y=0$ and stretched
along the $z$ direction.  The field of the vortex has
\begin{equation}
  \d_x\d_y \varphi- \d_y\d_x\varphi = 2\pi \delta(x)\delta(y) \,.
\end{equation}
The topological term is localized on the vortex core and has the form
\begin{equation}
  \L = -\pi\frac\rho m \delta(x)\delta(y) (u_x \dot u_y - u_y\dot u_x)
  + O(u^3) \,.
\end{equation}
Note that the leading term in this Lagrangian contains two powers of
$u$ but only one derivative, so it cannot be canceled by an unknown
interaction of the vortex core with the elastic waves.  Moreover,
other terms that couple $\theta$ and $X^a$ are expected to be
proportional to $\rho_s$ and are negligible.
\begin{figure}[tb]
%\vspace*{0.3cm}
\centerline{\includegraphics[width=7.5cm,angle=0]{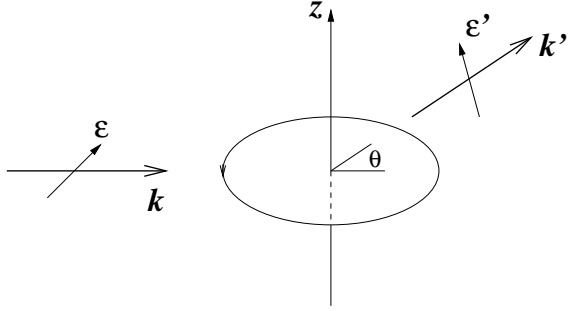}
}
\caption{\label{fig1}The scattering of transverse elastic phonons by a
superfluid vortex.}
\end{figure}

The process that will be considered has the following kinematics
(Fig.~\ref{fig1}).  A transverse phonon with momentum $\hbar\k$ and
linear polarization $\bm{\epsilon}$ falls perpendicularly onto a
vortex.  We are interested in the probability of its scattering into
the state with momentum $\hbar\k'$ and polarization $\bm{\epsilon'}$.
For simplicity, we assume that both $\bm{\epsilon}$ and
$\bm{\epsilon'}$ lie in the plane perpendicular to the vortex
and that the solid is isotropic.  The
matrix element of the process is
\begin{equation}
  M = \frac\pi m 
  (\bm{\epsilon}\times \bm{\epsilon'})\cdot\hat{\mathbf{z}}\,,
\end{equation}
from which we find the differential cross section per unit vortex
length,
\begin{equation}
  \frac{\d^2\sigma}{\d\theta\,\d l} = 
  \frac\pi2 \frac {\hbar^2} {m^2 v_\perp^2} k \sin^2\theta \,,
\end{equation}
where $v_\perp$ is the speed of transverse elastic waves.  This result
is a model-independent prediction of the effective Lagrangian
approach, valid at small $k$ and small superfluid fraction.
Qualitatively, the differential cross section has a linear dependence
on $k$ and is maximum when the scattering angle $\theta$ is
$90^\circ$.\vspace{6pt}

\emph{Conclusion.}---We have found the most general effective
Lagrangian describing the low-energy dynamics of supersolids.  We show
that, in the limit of small superfluid density, the Lagrangian
contains a topological term which has a fixed coefficient.  From this
term we computed the cross section of scattering of transverse
phonons off a superfluid vortex.  If the supersolid state is realized
in $^4$He, this prediction is, in principle, verifiable.

The formalism used in this paper can be extended to relativistic
systems, e.g., for describing the crystalline superfluid phases of
quark matter~\cite{Alford:2000ze}.  Instead of Galilean invariance,
one requires relativistic invariance of a theory of four Goldstone
bosons $\theta$ and $X^a$.  Such a theory, when coupled to gravity,
gives rise to a gravitational analog of the Anderson--Higgs mechanism.
Indeed, theories of this type have been proposed recently as an
infrared modification of gravity that gives the graviton
Lorentz-breaking mass terms~\cite{Rubakov:2004eb,Dubovsky:2004sg}.
From this point of view, the modification of gravity considered in
Refs.~\cite{Rubakov:2004eb,Dubovsky:2004sg} can be interpreted as an
effect coming from a supersolid dark matter sector.

The author thanks A.~Andreev, J.~Erlich, P.~Kovtun, M.\,A.~Stephanov,
and D.\,J.~Thouless for discussions.  This work was supported by DOE
grant DE-FG02-00ER41132 and the Alfred P.~Sloan Foundation.

\end{document}